\documentclass[twocolumn]{aastex62}

\usepackage{graphics,graphicx,xspace,natbib,amssymb}
\usepackage[caption=false]{subfig}
\usepackage{amsmath}
\usepackage{threeparttable}
\usepackage{makecell}

\usepackage{multirow}
\usepackage{fontawesome5}

\newcommand{\squiggle}{SQuIGG$\vec{L}$E\xspace}

\newcommand{\comment}[1]{}

\let\itAA\AA
\renewcommand{\AA}{\mathrm{\itAA}}

\shorttitle{HeavyMetal + ALMA}
\shortauthors{Suess et al.}

\begin{document}

\title{Cold gas in a post-starburst pair at $z\sim1.4$: major mergers as a pathway to quenching in the HeavyMetal survey}

\author[0000-0002-1714-1905]{Katherine A. Suess}
\affiliation{Department for Astrophysical \& Planetary Science, University of Colorado, Boulder, CO 80309, USA}

\author[0000-0002-9861-4515]{Aliza G. Beverage}
\affiliation{Department of Astronomy, University of California, Berkeley, CA 94720, USA}

\author[0000-0002-7613-9872]{Mariska Kriek}
\affiliation{Leiden Observatory, Leiden University, P.O. Box 9513, 2300 RA Leiden, The Netherlands}

\author[0000-0003-3256-5615]{Justin~S.~Spilker}
\affiliation{Department of Physics and Astronomy and George P. and Cynthia Woods Mitchell Institute for Fundamental Physics and Astronomy, Texas A\&M University, 4242 TAMU, College Station, TX 77843-4242, US}

\author[0000-0001-5063-8254]{Rachel Bezanson}
\affiliation{Department of Physics and Astronomy and PITT PACC, University of Pittsburgh, Pittsburgh, PA 15260, USA}

\author[0000-0002-1759-6205]{Vincenzo~R.~D'Onofrio}
\affiliation{Department of Physics and Astronomy and George P. and Cynthia Woods Mitchell Institute for Fundamental Physics and Astronomy, Texas A\&M University, 4242 TAMU, College Station, TX 77843-4242, US}

\author[0000-0002-5612-3427]{Jenny E. Greene}
\affiliation{Department of Astrophysical Sciences, Princeton University, 4 Ivy Lane, Princeton, NJ 08544, USA}

\author[0000-0002-3101-8348]{Jamie Lin}
\affiliation{Department of Physics and Astronomy, Tufts University, Medford, MA 02155, USA}

\author[0000-0002-0696-6952]{Yuanze~Luo}
\affiliation{Department of Physics and Astronomy and George P. and Cynthia Woods Mitchell Institute for Fundamental Physics and Astronomy, Texas A\&M University, 4242 TAMU, College Station, TX 77843-4242, US}

\author[0000-0002-7064-4309]{Desika Narayanan}
\affil{Department of Astronomy, University of Florida, 211 Bryant Space Science\
s Center, Gainesville, FL 32611 USA}
\affil{Cosmic Dawn Center at the Niels Bohr Institute, University of Copenhagen\
 and DTU-Space, Technical University of Denmark}

\author[0000-0002-7075-9931]{Imad Pasha}
\affiliation{Dragonfly Focused Research Organization, 150 Washington Avenue, Santa Fe, NM 87501, USA}
\affiliation{Department of Astronomy, Yale University, 219 Prospect Street, New Haven, CT 06511, USA}

\author[0000-0002-0108-4176]{Sedona H. Price}
\affiliation{Space Telescope Science Institute (STScI), 3700 San Martin Drive, Baltimore, MD 21218, USA}

\author[0000-0003-4075-7393]{David~J.~Setton}\thanks{Brinson Prize Fellow}
\affiliation{Department of Astrophysical Sciences, Princeton University, Princeton, NJ 08544, USA}

\author[0000-0003-1535-4277]{Margaret E. Verrico}
\affiliation{University of Illinois Urbana-Champaign Department of Astronomy, University of Illinois, 1002 W. Green St., Urbana, IL 61801, USA}
\affiliation{Center for AstroPhysical Surveys, National Center for Supercomputing Applications, 1205 West Clark Street, Urbana, IL 61801, USA}

\author[0000-0001-6454-1699]{Yunchong Zhang} 
\affiliation{Department of Physics and Astronomy and PITT PACC, University of Pittsburgh, Pittsburgh, PA 15260, USA}

\email{suess@colorado.edu}

\begin{abstract}
Recent observations at low redshift have revealed that some post-starburst galaxies retain significant molecular gas reservoirs despite low ongoing star formation rates, challenging theoretical predictions for galaxy quenching. To test whether this finding holds during the peak epoch of quenching, here we present ALMA CO(2–1) observations of five spectroscopically confirmed post-starburst galaxies at $z\sim1.4$ from the HeavyMetal survey. While four galaxies are undetected in CO emission, we detect $M_{\rm H_2}\sim10^{9.7}~M_\odot$ of molecular gas in one system. The detected system is a close pair of massive ($M_* = 10^{11.1-11.2}M_\odot$) post-starburst galaxies with no clear tidal features, likely caught in the early stages of a major merger. These results suggest that mergers may be a key factor in retaining molecular gas while simultaneously suppressing star formation in quenched galaxies at high redshift, possibly by driving increased turbulence that decreases star formation efficiency. Unlike previous studies at $z<1$, we find no correlation between molecular gas mass and time since quenching. This may be explained by the fact that --- despite having similar UVJ colors --- all galaxies in our sample have post-burst ages older than typical gas-rich quenched systems at low redshift.  
Our results highlight the importance of major mergers in shaping the cold gas content of quiescent galaxies during the peak epoch of quenching.
\end{abstract}


\section{Introduction}

One of the most intriguing puzzles in galaxy formation is the quiescent nature of massive early-type galaxies. There should be plenty of fuel for star formation in these galaxies: gas from the intergalactic
medium is continuously falling onto their dark matter halos, and their evolving stellar populations have substantial mass-loss rates. As a result, most cosmological simulations must resort to ad hoc models to heat or remove cold gas in order to halt the star formation in these massive galaxies. Feedback from active galactic nuclei (AGN) --- where quasar outflows expel gas from the galaxy, then the radio mode maintains quenching by heating the gas --- is one of the most popular of these models \citep[e.g.,][]{croton06,hopkins06}. This AGN feedback and subsequent quenching can be triggered by major mergers \citep[e.g.][]{springel05,wellons15}. Mechanisms that make cold gas accretion inefficient --- due to virial shocks \citep[e.g.,][]{dekel06} or the mass of the halo \citep[e.g.,][]{dave17,feldmann17} --- have also been proposed to quench massive galaxies. The one aspect that these models all have in common though, is that recently-quenched galaxies should not retain substantial cold molecular gas reservoirs.

In contrast to these theoretical predictions, observations of recently-quenched ``post-starburst" galaxies at $z<1$ have found that these quiescent galaxies can retain gas fractions of up to $\sim20\%$ --- similar to star-forming galaxies of similar mass and redshift, and an order of magnitude higher than expected for galaxies with such suppressed star formation rates \citep[e.g.,][]{french15,suess17}. Interestingly, the gas fraction in these galaxies appears to be a strong function of post-quenching age: the youngest post-starburst galaxies are often gas-rich, while galaxies which quenched $\gtrsim150$~Myr before observation are uniformly gas-poor \citep{french18,bezanson22}. Given the low star formation rates in these galaxies, it would take far longer than 150~Myr to consume all of the residual gas. 
These observations raise several puzzles: how are these galaxies able to retain such large gas reservoirs after they quench? Why does this gas disappear so quickly after quenching? And what do these observations tell us about the mechanisms responsible for quenching?

One key caveat of these $z<1$ observational studies is that they have targeted unusual galaxies: most quiescent galaxies stop forming stars at higher redshifts of $z\sim1-3$ \citep[e.g.,][]{whitaker12,wild16,belli19} and may quench via different physical mechanisms than their low-redshift counterparts. 
To understand the processes responsible for creating the bulk of the quiescent population, we need to measure the molecular gas properties of post-starburst galaxies during the peak quenching epoch and compare their star-formation efficiencies to typical star-forming galaxies at similar redshifts. 

To date, only a few studies have attempted to directly detect cold gas in $z>1$ quiescent galaxies. \citet{bezanson19} and \citet{williams21} placed limits of $f_{\rm{gas}}<2-7\%$ in seven massive quiescent galaxies at $z\sim1.5$; however, given the old stellar ages of these galaxies and the strong age-gas trend observed at lower redshift, these non-detections are unsurprising. \citet{belli21} finds $f_{\rm{gas}}\sim13-23\%$ in three young quiescent galaxies at $z\sim1$; however, these galaxies were specifically selected to have 24$\mu$m excesses, and may not be typical of the broader population of quenching galaxies. \citet{zanella23} targeted two slightly lower-mass young quiescent galaxies ($\log{M_*/M_\odot}\sim10.6$) and detected both, with gas fractions of $f_{\rm{gas}}\sim8-16\%$. Of these studies, only one of the \citet{zanella23} targets probes the youngest (light-weighted age $t_{50}\lesssim 200$Myr) post-starburst galaxies that $z<1$ studies imply may retain much larger gas reservoirs. In large part, this observational gap was due to historical difficulties spectroscopically-confirming quiescent galaxies at high redshifts: both redshifts and ages must be measured from relatively faint absorption features that are shifted into the rest-frame near-infrared at $z\gtrsim1$, which from the ground requires very long integrations with infrared spectrographs. JWST/NIRSpec has recently revolutionized this field, taking detailed spectra of dozens of quiescent galaxies from $1\lesssim z\lesssim 5$ \citep[e.g.,][]{carnall23,carnall25,glazebrook24,nanayakkara24,belli24,slob24,setton24,degraaff25,siegel25}; however, most of these objects have yet to be observed in cold gas with ALMA.

In this Letter, we present ALMA observations of the molecular gas contents of five young post-starburst galaxies at $z\sim1.4$. This work is enabled by ultradeep Keck/LRIS+MOSFIRE observations which allow us to precisely confirm spectroscopic redshifts and young post-quenching ages in our sample. We find that four of these young quiescent galaxies at $z\sim1.4$ are undetected. We significantly detect one object, which is part of an ongoing major merger of two massive post-starburst galaxies. These results suggest that major mergers may play a key role in triggering quenching while retaining molecular gas at these early cosmic times.

Throughout this Letter, we assume a cosmology of $\Omega_m = 0.3$, $\Omega_\Lambda = 0.7$, and $h = 0.7$; we quote AB magnitudes. 

\section{Data \& Methods}

The galaxy sample was drawn from the Heavy Metal survey \citep{kriek24}, which obtained ultra-deep spectra of 21 quiescent galaxies at $z\sim1.4$ and $z\sim2.1$ with LRIS and MOSFIRE on the Keck I Telescope. The Heavy Metal survey was executed in the COSMOS/UltraVISTA field \citep{mccracken12} and thus deep multiwavelength photometry and high-resolution imaging are also available for all galaxies \citep{muzzin13,scoville07,mowla19}. The galaxies span a range in stellar mass of ${\rm log} M_*/M_\odot = 10.8-11.9$. For more details on the survey design, sample selection, observations, data reduction and analysis, stellar population measurements, and sample characteristics, we refer to \citet{kriek24}. Additionally, in \citet{beverage23} we present the stellar population ages, chemical compositions, and velocity dispersions for the Heavy Metal quiescent galaxies.

\begin{figure}[ht]
    \centering
    \includegraphics[width=0.45\textwidth]{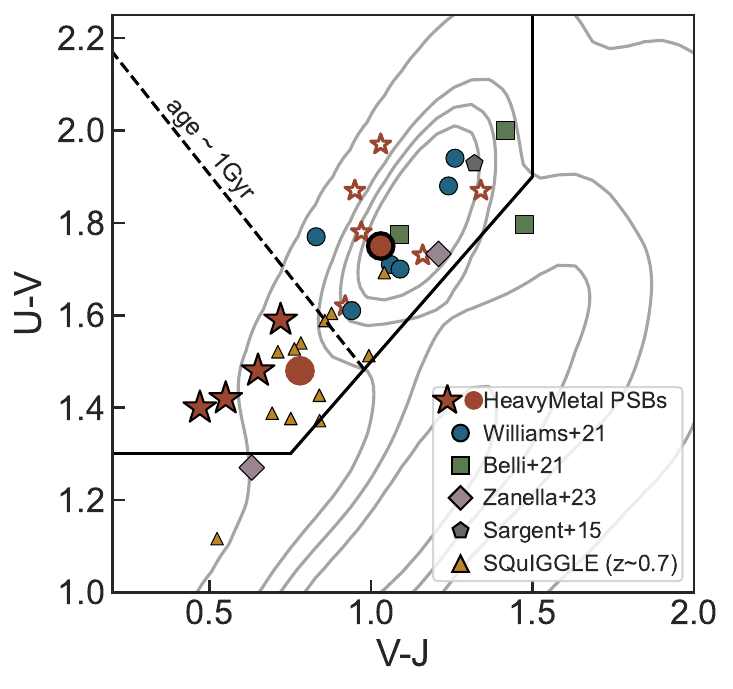}
     \caption{UVJ diagram for our sample of young post-starburst galaxies at $z\sim1.4$ (red stars; red circle for HM1-214340 and red circle with black outline for serendipitously-observed UVISTA-214345) as well as existing measurements of molecular gas in quiescent galaxies at $z>1$ \citep{williams21,bezanson19,belli21,sargent15,zanella23} and $z\sim0.7$ (\squiggle; \citealt{bezanson22}). UVJ colors for \squiggle galaxies are extrapolated from their best-fit SED models \citep{suess22}. The grey contours show the distribution of $1<z<1.5$ galaxies from 3D-HST \citep{skelton14}. With the exception of one of the \citet{zanella23} objects, most previous measurements of gas in quiescent galaxies at $z>1$ have targeted older systems ($\gtrsim1$Gyr old from the \citealt{belli19} relation). Our sample (filled red stars; selected from the full HeavyMetal sample in open red stars) focuses on the youngest quiescent galaxies, where most existing measurements are at $z<1$ \citep{bezanson22}. }
    \label{fig:1}
\end{figure}

\subsection{ALMA observations}
\label{sec:alma}

\begin{deluxetable*}{cccccccccc}[!t]
\tabletypesize{\scriptsize}
\tablecaption{Properties of the ALMA observations}
\tablehead{
    \colhead{ID} & \colhead{RA} & \colhead{Dec} & \colhead{$z_{\rm{spec}}$} & \colhead{Integration Time} & \colhead{Angular Resolution} & \colhead{Continuum Flux} & \colhead{SdvCO(2--1)} & \colhead{L$^\prime_{\rm{CO(2-1)}}$} & \colhead{$\log{\rm{M}_{H_2}/\rm{M}_\odot}$}\\
    \colhead{} & \colhead{[hh:mm:ss]} & \colhead{[dd:mm:ss]} & \colhead{} & \colhead{[s]} & \colhead{[$\arcsec$]} & \colhead{[$\mu$Jy]} & \colhead{Jy km/s} & \colhead{$10^9$K km/s pc$^2$} & \colhead{}
}
\startdata
HM1-214340 & 10:00:37.094 & 02.34.10.284 & 1.418 & 8437 & 2.418 & $<14.43$ & $0.05\pm0.01$ & $1.38 \pm 0.28$ & $9.74 \pm 0.09$\\
HM2-23351 & 10:00:48.4350 & 01.43.24.740 & 1.421 & 5625 & 2.436 & $<18.00$ & $<0.04$ & $<0.95$ & $<9.58$ \\
HM2-25702 & 10:00:49.197 & 01.45.27.470 & 1.419 & 8437 & 1.536 & $<18.91$ & $<0.06$ & $<1.55$ & $<9.79$\\
HM2-213947 & 10:00:50.683 & 02.33.55.368 & 1.397  & 8346 & 1.523 & $<19.29$ & $<0.04$ & $<1.02$ & $<9.61$\\
HM1-217249 & 10:00:45.622 & 02.36.19.297 & 1.377 & 7530 & 1.433 & $<18.62$ & $<0.10$ & $<2.55$ & $<10.01$ 
\enddata
\tablecomments{Upper limits for the undetected sources are 3$\sigma$, and assume an 800km s$^{-1}$ line width.}
\label{table:alma}
\end{deluxetable*}
    
To understand the role of molecular gas in the quenching of star formation, we observed the five youngest $z\sim1.4$ quiescent galaxies from the HeavyMetal survey with ALMA. These galaxies were selected for ALMA observations by their rest-frame UVJ colors, as shown in Figure~\ref{fig:1} \citep[e.g.,][]{williams09,whitaker12_psb,belli19}. The galaxies have stellar masses around $\sim 10^{11} M_\odot$ and show multiple detected stellar absorption lines in their spectra. They all have wavelength coverage of the H$\alpha$ emission line, from which we derived (limits on) their star formation rates, as shown in Table 3 of \citet{kriek24}. 

For all five galaxies we targeted CO(2–1) in ALMA Band 3 as part of 2019.1.01286.S (PI Kriek). Observations were made in three ${\sim}80$-minute blocks, with total integration times ranging from 1.5-2.3hr, and angular resolution $\sim1.4-2.4"$ as shown in Table~\ref{table:alma}. Data were reduced using the standard ALMA reduction pipeline. 

We created $\sim2$mm continuum images using the full bandwidth of our observations (excluding the 500 km s$^{-1}$ around the CO(2-1) line). No continuum emission was detected in any of our primary targets, either at the naturally-weighted image or in a UV-tapered image (3$\sigma$ limiting fluxes $\lesssim20\mu$Jy; see Table~\ref{table:alma}). These continuum nondetections are consistent with the low ongoing SFRs inferred from H$\alpha$ by \citet{kriek24}. 

We extract CO(2-1) spectra for all targets at an effective velocity resolution of $\sim$200 km s$^{-1}$. We detect CO(2-1) emission at $\sim5\sigma$ in HM1-214340; no other targets are detected. For the four undetected sources, we extract spectra using CASA's \texttt{imfit} task, fixing the centroid to the phase center and the size to 2". Following \citet{bezanson22}, we conservatively estimate the upper limit on the CO(2-1) line flux using a 800 km s$^{-1}$ channel and report values in Table~\ref{table:alma}. While HM1-214340 is clearly detected, we do not observe any evidence for velocity gradients or resolved structure at the depth and resolution of our observations. Therefore, we first fit for the source position and size in a single $\sim$1000 km s$^{-1}$ channel, then fix the position and size in \texttt{imfit} to extract a spectrum at $\sim$200 km s$^{-1}$ velocity resolution, allowing only the intensity to vary channel-to-channel. We fit a single gaussian to the spectrum to find the integrated CO(2-1) line flux and its uncertainty. For all objects we convert (limits on) CO(2-1) line flux to molecular gas mass assuming thermalized emission \citep[e.g., $r_{21}=1.0$,][]{combes07,dannerbauer09,young11} and a Milky Way-like $\alpha_{CO}=4.0$ (see \citealt{bolatto13} and references therein, as well as discussions in \citealt{bezanson22} and \citealt{spilker22}).

\subsection{Spectral modeling}
One of our goals is to test whether the correlation between gas mass and time since quenching identified in \citet{bezanson22} (see also \citealt{french18}) holds at higher redshift. This test requires that we calculate galaxy ages that are directly comparable to those used by \citet{bezanson22}, which use the \texttt{Prospector} modeling framework described in detail in \citet{suess22}.

We use \texttt{Prospector}~v2 \citep{johnson20} to simultaneously fit UltraVISTA photometry \citep{muzzin13} and Keck/LRIS-R, Keck/MOSFIRE-J, and Keck/MOSFIRE-H spectroscopy \citep{kriek24}. We calibrate each spectrum separately using a fifth-order polynomial. We adopt a 5\% error floor for all photometric data points. Our model assumes a \citet{chabrier03} IMF, the \citet{kriek13} dust attenuation law with free dust index and normalization (with young stars fixed to be twice as attenuated as old stars), and the non-parametric post-starburst star-formation history from \citet{suess22}. We include nebular emission, tying the gas-phase metallicity to the stellar metallicity and treating the ionization parameter as a free parameter. All parameters use uniform priors, except for stellar metallicity, which follows a prior approximating the stellar mass–metallicity relation of \citet{gallazzi05}. We generate stellar populations using FSPS \citep{conroy09,conroy10}, adopting the MILES spectral library \citep{falcon11} and MIST isochrones \citep{dotter16,choi16}, computed with MESA \citep{paxton11}. Posterior distributions are sampled using the dynamic nested sampling algorithm implemented in \texttt{dynesty} \citep{speagle20}.

\section{Results}
\label{sec:results}

\begin{figure*}[ht]
    \centering
    \includegraphics[width=0.95\textwidth]{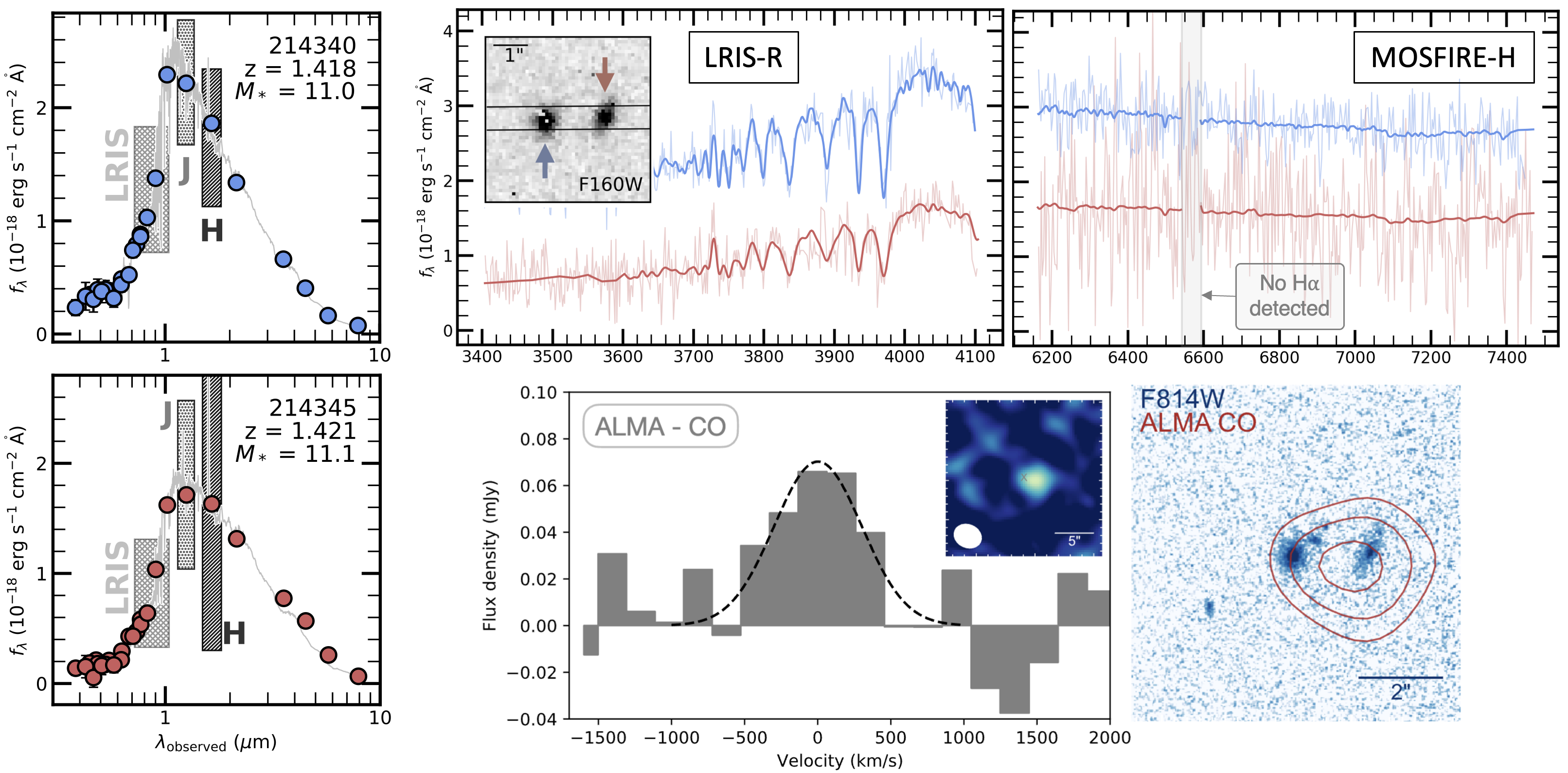}
    \caption{Top center/right: LRIS and MOSFIRE spectra of UVISTA-214340 (blue) and UVISTA-214345 (red). Left: SEDs and best-fit \texttt{Prospector} models to UVISTA-214340 and UVISTA-214345. Bottom center: CO(2-1) spectrum and moment 0 map of our detection. Bottom right: ALMA contours (3$\sigma$, 4$\sigma$, 5$\sigma$) overlaid on the F814W image of the region. Both galaxies are spectroscopically confirmed massive post-starburst galaxies at the same redshift, likely in the early stages of a near-equal-mass merger. The ALMA observations show $\sim10^{9.74}\rm{M}_*$ of molecular gas contained in the system; as described in the text, the majority of this gas is likely associated with UVISTA-214345, but our spatial resolution is too coarse to conclusively determine whether some gas is also associated with HM1-214340 or one of the small clumps apparent in F814W.}
    \label{fig:detection}
\end{figure*}

We significantly detect ($\sim5\sigma$) molecular gas in one of our five primary targets, HM1-214340. Figure~\ref{fig:detection} shows the ALMA CO spectrum and moment-0 map as well as the best-fit \texttt{Prospector} model to the photometry and Keck spectroscopy. Remarkably, our Keck observations serendipitously obtained a spectrum of UVISTA-214345, offset just $\sim1\farcs76$ ($\sim15$~kpc) to the west of HM1-214340. The combination of the UltraVista photometry and Keck spectroscopy confirm that {\it both} HM1-214340 and UVISTA-214345 are massive ($10^{11.0}$ and $10^{11.1}M_\odot$) post-starburst galaxies at very similar redshifts: at $z=1.418$ and $z=1.421$, the two systems are separated by just $\sim300$km/s. At the depth of existing HST/F814W and F160W imaging, no tidal features or significant asymmetries are present; however, as discussed in more detail below, there are several small offset clumps in the F814W image that may be associated with the system. Altogether, the imaging suggests that HM1-214340 and UVISTA-214345 are in the early stages of a near 1:1 major galaxy merger: they may or may not have had one or more close passes, but have not yet reached coalescence. No 2mm continuum emission is detected from the system at the depth of our observations.

While the angular resolution of the ALMA data ($\sim2.4"$; Table~\ref{table:alma}) is larger than the offset between the two galaxies, the data are consistent with the western source UVISTA-214345 being the primary source of the CO flux. We first fit a single 2D gaussian to the moment-0 map of the CO flux, allowing all fit parameters including the position to be free. The best-fit CO position is (10:00:36.99, +02:34:10.0) with $\sim0\farcs5$ uncertainty, consistent with the position of UVISTA-214345. As discussed in Section~\ref{sec:alma}, the CO emission is consistent with a point source at the resolution of our data. We then model the CO flux with two 2D gaussians, one fixed to the position of UVISTA-214345 and one fixed to the position of HM1-214340, fixing each gaussian to the size and position angle of the clean beam and leaving only the amplitude of each component free. In this two component fit, we find the best-fit flux for UVISTA-214345 is $0.105\pm0.021$~Jy~km~s$^{-1}$ and the best-fit flux for HM1-214340 is $0.031\pm0.021$~Jy~km~s$^{-1}$. We note that the sum of these two fluxes is in good agreement with the flux in the single-component fit. This fit indicates that $\gtrsim80\%$ of the CO flux is likely originating from UVISTA-214345, while HM1-214340 could contribute up to $\sim25\%$ of the CO flux but is not significantly detected. We additionally note that the HST/F814W imaging, shown in the bottom right of Figure~\ref{fig:detection}, shows several small offset clumps to the N/NW of both HM1-214340 and UVISTA-214345. These clumps are not visible in the HST/F160W imaging, which could either indicate that the clumps are relatively blue and associated with recent star formation, at a different redshift than our target system, or simply too faint for the shallow F160W imaging. While it is also possible the CO emission is partially associated with one of these clumps, the likeliest solution is that the CO emission is primarily coming from UVISTA-214345. 
Table~\ref{table:alma} reports the CO line flux for the HM1-214340 / UVISTA-214345 system from the single-component fit above. This line flux corresponds to $10^{9.74}M_\odot$ of molecular gas, or a gas fraction of $\sim4$\%.

\begin{figure*}[ht]
    \centering
    \includegraphics[width=0.95\textwidth]{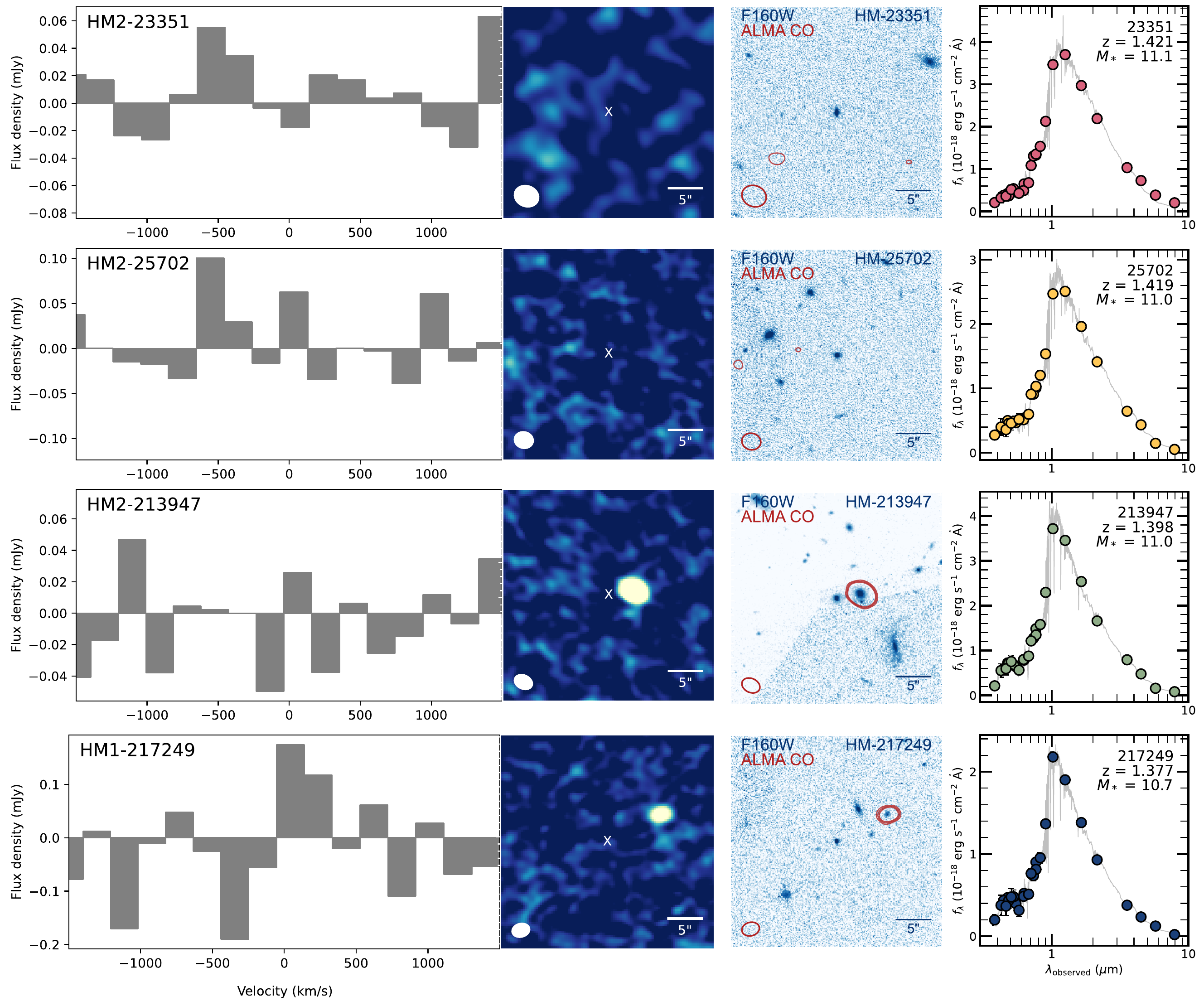}
    \caption{Non-detections of our primary HeavyMetal targets. Left: ALMA CO(2-1) spectra extracted at the position and redshift of our target. Center: CO(2-1) line map; white ``x" at the center shows the position of our target. Right: HST-F160W images with CO contours overlaid (3$\sigma$, 4$\sigma$, 5$\sigma$). No sources are detected in the cubes for HM2-23351 or HM2-57702. In the HM-213947 cube we serendipitously detect UVISTA-214044 at $z\sim1.4$; in the HM1-217249 cube we serendipitously detect 217299 at $z\sim1.38$. Right: multi-band photometry and \texttt{Prospector} fits to each galaxy. }
    \label{fig:nondetections}
\end{figure*}

Our other four primary targets are not detected in either CO(2-1) or 2mm continuum, as shown in Figure~\ref{fig:nondetections}. This allows us to place $3\sigma$ upper limits on their molecular gas mass of $\lesssim10^{9.5-10.0}M_\odot$ (see Table~\ref{table:alma}). 
We note that while our primary targets HM2-213947 and HM1-217249 are not detected, both have nearby neighbors that are significantly detected in CO(2-1), as shown in the right columns of Figure~\ref{fig:nondetections}. In the HM2-213947 cube we serendipitously detect UVISTA-214044 at $z\sim1.4$ and a projected separation of $\sim30$kpc; in the HM1-217249 cube we serendipitously detect 217299 at $z\sim1.38$ and a projected separation of $\sim70$kpc. Given the similar redshifts and close on-sky separations of these gas-rich galaxies from our primary targets, we infer that they are likely part of close pairs (with larger separations than the HM1-214340/UVISTA-214345 system) or the same overdense environment as our primary targets.

\begin{figure*}
    \centering
    \includegraphics[width=0.98\textwidth]{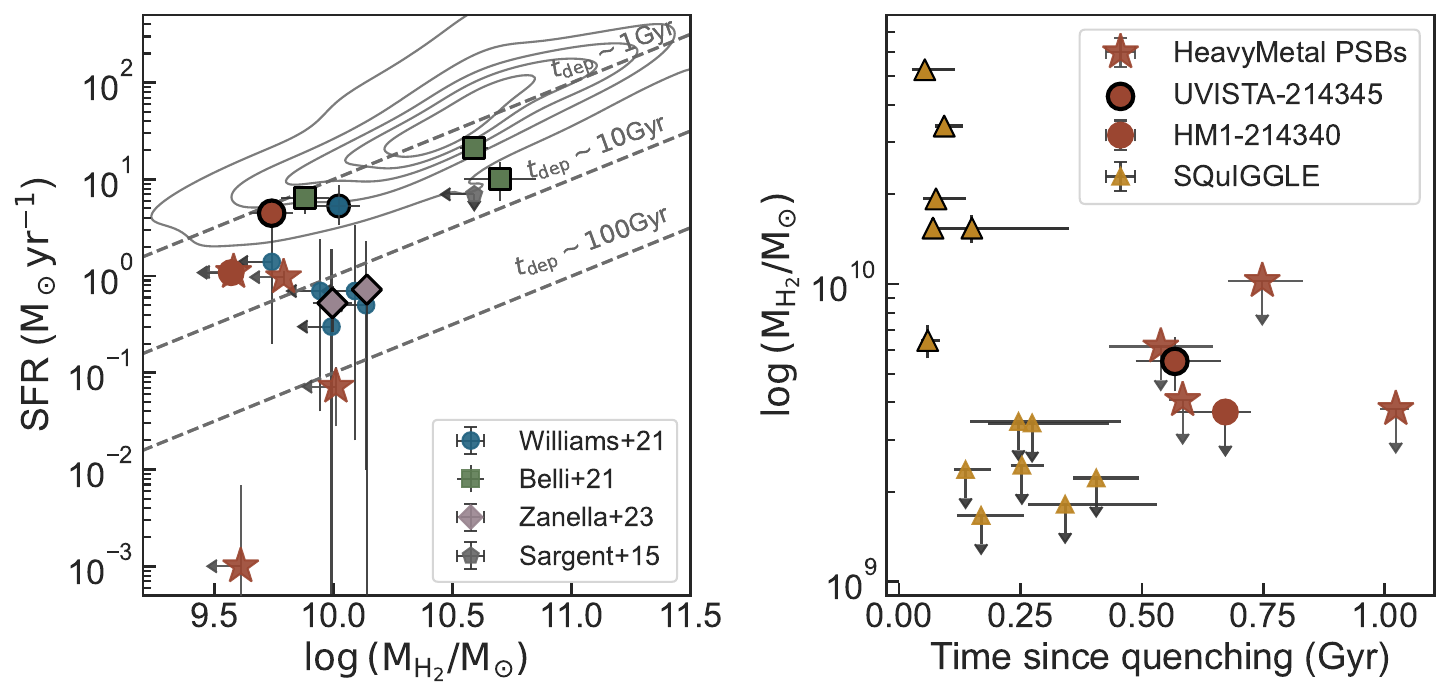}
    \caption{Left: star formation rate as a function of molecular gas mass for a sample of star-forming galaxies at $z\sim0.5-1.5$ (PHIBSS \& PHIBSS-2; \citealt{tacconi13,freundlich19}; grey contours) and quiescent galaxies at $z>1$ (colored points). SFRs for our sample are from our \texttt{Prospector} fits, while SFRs for literature samples are taken from the relevant paper. Gas detections are outlined in black. Dashed lines indicate constant depletion times ($t_{\rm{dep}}\equiv M_{H_2}/\rm{SFR}$). Our observations are at or below the expectations for star-forming galaxies, with depletion times longer than 1 Gyr. Right: molecular gas mass as a function of time since quenching both for our sample of PSBs at $z\sim1.4$ and the \squiggle sample of PSBs at $z\sim0.7$ \citep{bezanson22}. While the \squiggle galaxies show a clear trend that gas-rich galaxies quenched more recently, no clear trend is present in our observations at higher redshift.}
    \label{fig:analysis}
\end{figure*}

In Figure~\ref{fig:analysis}, we place our sample in the context of existing observations of gas in quiescent galaxies at $z>0.5$. The left panel shows star formation rate as a function of molecular gas mass for all $z>1$ quiescent galaxies with molecular gas measurements from \citet{williams21,belli21,zanella23,sargent15}. Objects with detected gas are outlined in black to differentiate them from upper limits. We note that we associate our CO(2-1) detection with UVISTA-214345 (see Section~\ref{sec:results}), and place HM1-214340 at the 3$\sigma$ upper limit implied by our two-component modeling. Our detection has an inferred depletion time $\sim1$Gyr, in good agreement with the bulk of the PHIBSS sample shown in grey contours \citep[][]{tacconi13,freundlich19}. Our nondetected galaxies have upper limits on their depletion times, consistent with long depletion times up to $10-100$Gyr.

The right panel of Figure~\ref{fig:analysis} shows molecular gas mass as a function of time since quenching \citep[defined following][]{suess22}. Galaxies from the \squiggle survey \citep{suess22_squiggle,bezanson22} at $z\sim0.7$ are shown in yellow, while our sample at $z\sim1.4$ is shown in red. Again, we outline all gas detections in black. Interestingly, the \texttt{Prospector} fits yield uniformly older $t_{\rm{quench}}$ ages for HeavyMetal than for \squiggle, despite the fact that the UVJ colors of the two samples are consistent (Fig~\ref{fig:1}), indicating similar $t_{50}$ ages. While the \squiggle sample shows a clear trend that gas-rich galaxies have $t_{\rm{quench}}\lesssim150$~Myr and gas-poor galaxies have $t_{\rm{quench}}\gtrsim150$~Myr, all galaxies in our HeavyMetal sample have $t_{\rm{quench}}\gtrsim500$~Myr. No trend with age is apparent in the HeavyMetal sample: the detected system has similar $t_{\rm{quench}}$ as our non-detected systems.

\section{Discussion \& Conclusions}

In this Letter, we observed CO(2-1) as a tracer of molecular gas in five $z\sim1.4$ post-starburst galaxies with ultradeep spectroscopy from the HeavyMetal program. One of the five targets was detected, with an inferred molecular gas mass of $10^{9.74}M_\odot$; the other four targets were undetected in both CO(2-1) line emission and 2mm continuum. Our detected target HM1-214340 is in a close pair with UVISTA-214345, which we spectroscopically confirm to be a similarly-massive post-starburst galaxy at very similar redshift ($\Delta z=0.003$). No clear tidal features are detected and the galaxies have a projected separation of just $\sim15$kpc, implying that the two galaxies may be in the early stages of a major merger event. For two galaxies that were undetected in CO(2-1), HM2-213947 and HM1-217249, our ALMA data spectroscopically confirm nearby neighbors to both objects, with projected separations of $\sim30$ and $\sim70$~kpc, respectively. The remaining two undetected galaxies, HM2-23351 and HM2-25702, appear isolated in our ALMA cubes, though we note that the large number of massive quiescent galaxies in the HeavyMetal survey at $z\sim1.4$ indicates that all of the galaxies in this paper may preferentially live in an overdense environment. 

The strong correlation between molecular gas mass and post-burst age observed both at $z\sim0$ \citep{french18} and $z\sim0.7$ \citep{bezanson22} is not clearly reproduced in our $z\sim1.4$ sample. While our four non-detections are consistent with the lower-redshift trend -- galaxies with $t_{\rm{quench}}\gtrsim150$Myr lack substantial gas reservoirs -- our detected system is an outlier: a relatively older system ($t_{\rm{quench}}\sim600$Myr) that retains $\sim10^{9.7}~M_\odot$ of molecular gas. Interestingly, recent work directly studying the HI (not CO) reservoirs of low-redshift post-starburst galaxies finds no significant trend between post-burst age and gas mass \citep{ellison25}. Observations at a wider age range, particularly probing the youngest quiescent galaxies, are required to determine whether a trend between age and molecular gas mass persists at $z\gtrsim1$. Our detection of gas in an older system may also point to a distinct physical driver governing gas survival: perhaps {\it mergers}, not post-burst age, are the most important driver of whether a quenched galaxy is able to retain molecular gas. Merger signatures are more common for younger post-starburst galaxies \citep[e.g.,][]{verrico23,ellison24}, and this correlation could partially explain the gas-age trend observed at low redshift. 

In broad brush strokes, it is perhaps unsurprising that major mergers could play a key role in quenching and the retention of molecular gas in quiescent galaxies at $z\sim1.4$. Simulations have long suggested that mergers and quenching could be causally related, typically because mergers in the simulations are able to trigger enhanced star formation, black hole accretion, and subsequent AGN feedback and removal of molecular gas \citep[e.g.,][]{springel05,hopkins08,lotz21}. Observations have largely corroborated this picture: post-starburst galaxies have high rates of disturbed morphologies indicating recent mergers \citep[e.g.][]{pawlik18,sazonova21,wilkinson22,ellison22,verrico23,wu23}. However, the details of the HM1-214340/UVISTA-214345 system differ from this classical picture: we observe a close pair where both galaxies have already quenched, not a single post-coalescence quenched remnant. This suppression of star formation in a close pair is in contrast with previous studies that typically find star-formation is {\it enhanced} in the close pair stage \citep[e.g.,][]{barton2000,patton13,ferreira25}, with post-starburst signatures peaking $\sim500$Myr after coalescence \citep{ellison24}. 

Our gas-rich post-starburst pair may instead represent a system that already experienced first passage and an associated starburst before the time of observation, consistent with our modeled SFH where both galaxies experienced a recent starburst which ended ${\sim}500-600$Myr ago (Fig~\ref{fig:analysis}), and the current ${\sim}15$~kpc on-sky separation. In this picture, turbulence \citep[e.g.,][D'Onofrio et al. in prep.]{vandevoort18,smercina22,spilker22} may have temporarily prevented the molecular gas in the system from forming stars, leading to both post-starburst stellar populations and detectable CO. While stellar feedback from the starburst may also have played a role in decreasing star formation efficiency in this system, it typically acts on shorter timescales (see, e.g., discussion in \citealt{french21}). As the merger progresses, star formation may re-ignite and consume the remaining molecular gas, eventually leading to a gas-poor post-starburst remnant as in \citet{ellison24}, or perhaps the merger will remove the molecular gas from the system and prevent further star formation at all \citep[e.g.,][D'Onofrio et al. in prep.]{spilker22}. However, we cannot rule out the possibility that the process(es) that quenched HM1-214340 and UVISTA-214345 are entirely unrelated to their close separation.

Our spectroscopic analysis of these galaxies emphasizes discrepancies in different age indicators used for quiescent galaxies. Figure~\ref{fig:1} shows that the \squiggle and HeavyMetal samples have broadly consistent UVJ colors. Previous work indicated that these similar UVJ colors imply they have similar light-weighted ages \citep[$t_{50}$, e.g.,][]{whitaker12_psb,belli19,cheng25}. However, modeling the spectra with \texttt{Prospector} yields systematically older mass-weighted post-burst ages \citep[$t_{\rm{quench}}$, e.g.][]{suess22} for the HeavyMetal sample than the \squiggle sample. The difference between the average $t_{\rm{quench}}$ values for \squiggle and HeavyMetal ($\sim400$Myr) is significantly larger than the typical error bars on the ages ($\lesssim100$Myr). A detailed comparison of the spectra suggests that this difference in age between the two samples is reasonable: typical \squiggle spectra \citep{suess22_squiggle} have larger asymmetries between their calcium H \& K lines than our HeavyMetal sample, indicating younger ages because the calcium K line is blended with a deeper Balmer absorption feature.  This discrepancy between UVJ color-based ages, $t_{50}$, and \texttt{Prospector}-inferred ages, $t_{\rm{quench}}$, may arise from differences in the SFHs of the two samples: \squiggle galaxies are at lower redshift, and have a combination of both young and old stars \citep[A/(A+K)$\sim0.6$,][]{suess22_squiggle}. The presence of these older stars reddens the $V-J$ colors and increases $t_{50}$. Meanwhile, the HeavyMetal galaxies are at higher redshifts and have a smaller fraction of old stars, making their $t_{\rm{50}}$ younger for the same $t_{\rm{quench}}$. But even at fixed redshift, UVJ $t_{50}$ and \texttt{Prospector} $t_{\rm{quench}}$ can differ: UVISTA-214345 has a younger $t_{\rm{quench}}$ than HM1-214340 (Fig~\ref{fig:analysis}) but is located further up the UVJ quiescent sequence (Fig~\ref{fig:1}) nominally indicating older $t_{50}$. In this case, the UVJ colors are likely reddened by moderate dust obscuration: \texttt{Prospector} indicates UVISTA-214345 has A$_v\sim0.4$~mag, while HM1-214340 has just A$_v\sim0.07$~mag. If dusty post-starburst galaxies are more common at higher redshift \citep[e.g.,][]{setton24}, this may suggest that UVJ location alone is an imprecise age indicator for high-redshift quiescent galaxies \citep[see also][]{cheng25,antwi-danso23}. 
While differences between $t_{50}$ and $t_{\rm{quench}}$ are not necessarily unexpected, the fact that they yield wholly different conclusions --- either the HeavyMetal galaxies are the same age as \squiggle galaxies, or they are systematically older --- suggests that caution and careful methodology are required when computing and comparing the ages of quiescent galaxy samples, especially across different redshift ranges.

Our observations increase the total number of molecular gas measurements in quiescent galaxies at $z>1$ by a factor of $\sim$1.5 and emphasize the diversity of gas contents in these quenched systems; however, the physical mechanisms which quenched these galaxies and allow some but not all of them to retain gas remain unclear. 
While higher-resolution ALMA imaging that could pinpoint the spatial distribution of the molecular gas is likely out of reach for faint systems like HM1-214340/UVISTA-214345, deeper imaging with HST or JWST may reveal low surface-brightness tidal features and allow for a more extensive characterization of the merging system. Future observations of a larger sample of post-starburst galaxies may also allow us to disentangle the roles of merger history, environment, and internal feedback in the quenching process at cosmic noon. JWST spectroscopy offers a transformative path forward in terms of sample selection: while the HeavyMetal survey required nearly 100hr on-sky with Keck/MOSFIRE+LRIS to obtain ultradeep spectra of 20 massive quiescent galaxies, JWST/NIRSpec surveys like SUSPENSE \citep{slob24} can achieve similar multiplexing and higher SNR in dramatically shorter integration times of $\sim16$hrs, while surveys like EXCELS \citep{carnall24} extend our redshift lever arm to the beginning of the quenching epoch. These JWST/NIRSpec data will enable robust spectroscopic confirmation and age-dating of larger samples of post-starburst galaxies. If paired with additional deep ALMA observations, these data could unlock the ability to trace how molecular gas content evolves with time, environment, and merger history across cosmic noon.

\acknowledgements 
KAS, JSS, and VRD gratefully acknowledge support from NSF-AAG\#2407954 and 2407955, and NRAO-SOSPA7-016. This paper makes use of the following ALMA data: ADS/JAO.ALMA \#2019.1.01286.S, ADS/JAO.ALMA \#2016.1.01126.S and ADS/JAO.ALMA \#2017.1.01109.S. ALMA is a partnership of ESO (representing its member states), NSF (USA) and NINS (Japan), together with NRC (Canada), MOST and ASIAA (Taiwan), and KASI (Republic of Korea), in cooperation with the Republic of Chile. The Joint ALMA Observatory is operated by ESO, AUI/NRAO and NAOJ. The National Radio Astronomy Observatory is a facility of the National Science Foundation operated under cooperative agreement by Associated Universities, Inc. 

\software{astropy \citep{astropy2013, astropy2018},  
          Seaborn \citep{waskom17},
          }

\bibliographystyle{aasjournal}
\bibliography{all}

\end{document}